New additional material of meteor showers during 9$^{th}$ -19$^{th}$ centuries in the Islamic history.

Hassan M. Basurah
Astronomy Department, Faculty of Science
King-Abdul Aziz University, Jeddah, Saudi Arabia

E-Mail: hbasurah@kau.edu.sa

**Abstract**

This article presents twelve records of meteor showers in Arabic chronicles covering period from the 9th to the 19th century. The observations were in Egypt, Morocco, Syria and Yemen. These new addition historical records are considered to be important events which indicate a serious current interest in astronomy.

**Introduction:**

No doubts that the historical records of meteor showers are expected to be fruitful studying the evaluation of the solar system. The meteor showers at European records have studied by Dall'olmo (1978), at Japanese by Imoto and Hesagawa (1958) and Hesagawa (1993), at Chinese by Zhuang T. (1966), and at Arabia by Rada and Stephenson (1992) and Cook (1999).

The investigated material of meteor showers during the medieval are quite important in various aspects of astronomical research such as studying relationship between the comet duration and the peak of the zenithal hourly rate; Brown (1999), or study the orbital elements of their parent bodies and predicted radiant points; Hasegawa (1999), the distribution of dust surrounding comets such as Tempel-Tuttle; Yeomans (1981), periodicity of meteoric events, Rasmussn (1990, 1991), and extra.

Here is a collection of more extra meteor showers events from Arabic chronicles which are not appeared at the previous works. Searching of these materials was not within astronomical chronicles but through historical chronicles from different Islamic countries such as Egypt, Morocco, Syria and Yemen.

The date which used for these recording was Islamic Hijri calendar. The names of the Islamic months in sequence are Muharram, Safar, Rabi I, Rabi II, Jumada I, Jumada II, Rajab, Shaban, Ramadan, Shawwal, Thu-alquda, and Thu-alhijja. The Islamic Hijri calendar dates have been converted to the Julian calendar. Eight of these recorder dates are specified in full: year, lunar month, and day, while the other four annals give just the month or the year of occurrence, so the dates of such events have ranged between the beginning and the ending of the Islamic year or month.

**The historical events and interpretations:**

Descriptions of the 12 meteor showers are given at table 1 with its equivalent Julian dates, place of occurrence and references. Event no.1 matches the record of Korean meteor shower at Autumn 848, Hasegawa (1999). Record no. 9 is for the well known Leoind meteor shower at 1866. Events no. 10 of Nov. 1872 is identified with recorded one at Japan Hasegawa (1999).

**More identifications and corrections for the previous catalogs:**

There are 2 catalogs for the meteor showers, Rada and Stephenson (1992) and Cook (1999) and there are few points about them need to have attention about them.

1- The date of meteor shower of 21\1\1060 [15\12\451H] at Cook (1999) was not the correct date, it is 13\ 12\ 1059 [6\11\451H]. Cook mention a possibility of an error of using Christian dates, but it is clearly written as in the middle of Dhu alhijja .

2- The correct date of meteor shower of Reda (1992) at August 1197 [594H] is the end of 1199 [the beginning of 596H].

3- The event of August/September 1026 [Rajab 417H] at Reda (1992), was not meteor shower but a meteoroid at October / November 1026 [Ramadan 417H].

4- There was no meteor shower in 952 [341H] as appeared at Reda (1992), it was the well known meteor shower of 16 Oct. 855 [241H].

Conclusion:

Twelve events were collected from different Arabian literatures, which consider to be a new additional material of meteor showers. This will fill in the gaps for the early Muslim period and provide some extra corroborating material, which could be utilized for the purpose of calculation or improving orbital parameters.

**Table (1): The list of new material of the comets and meteors from different Arabian literatures, Islamic dates, country, and references.**

1- During the interval time between 16 August 847 to 4 August 848 [year 233H], a storm developed in Baghdad, the entire horizon became black, from afternoon till sunset. Stars hammered down from sunset till dawn.

                     Syria  [3, 7, 8]

2- On Tuesday 27 July 1377 [20 Rabi I 779 H], stars fell down at night. Peoples got a great horrified of it.

                     Egypt  [2, 9]

3- On Thursday night of 24 September 1532 [24 Safar 939 H], planetoids fell down before dawn and the sign continued to the limit that we never known except beyond our times.

                     Yemen  [7]

4- During late at night of 27 September 1566 [13 Rabi I 974H], lots of intensely luminous stars scattered right and left from the direction of Mecca (north - west).

                     Yemen  [7]

5- On Tuesday late at night of 7 November 1605 [26 Jumada II 1014H], a proliferation of stars occurred at the West of Ursa Major to Northwest. The entire ground lighted up and the stars appeared very chaotic to the point that everyone saw them was confused. That proliferation continued for an hour.

                     Yemen  [11]

6- During the interval time between 24 December 1650 to 12 December 1651 [year 1061H], late at night a proliferation of stars occurred. The ones who saw that got really scared.

                     Yemen  [1]

7-     At Nov. 1806 [Ramadan 1221H], stars flew east and west, and every one talks about it.

                           Yemen  [10]

8-     At 17 January 1837 [10 Shawwal 1252H], meteors got mixed up from Aden (south) to direction of Mecca (north) and from the north to the south for an hour at midnight.

                           Yemen  [6]

9-     On Tuesday night of 12 November 1866 [5 Rajab 1283H], just before dawn the stars proliferated in all directions and continued that way to point that no one could have counted them. Only the light of the daybreak covered that up.

                           Yemen  [7]

10-    In November 1872 [Ramadan 1289H], the stars proliferated from the beginning of the night until dawn. People thought it was Judgment Day.

                           Yemen  [5]

11  -  On 14 or 15 October 1874  [4 or 5 Ramadan 1291H], an occurrence of planetoids with great and troubling chaos where some of them were coming from the east, others coming from the west while others from other directions. That continued until late at night.

                           Morocco  [4]

12-    On 26 November 1885 [19 Safar 1303H], the stars were very dispersed in every direction and fell down and other things contrary to what's normal.

                           Morocco  [4]